\documentclass[10pt,twocolumn,twoside,submit]{JCNtran}

\usepackage[bookmarks=true]{hyperref}
\usepackage{colortbl}
\usepackage{pdflscape}
\usepackage{epstopdf}
\usepackage{epsfig}
\usepackage{standalone}
\usepackage{amssymb}
\usepackage{subfigure}
\usepackage{amsthm,amsmath}
\usepackage{float}
\usepackage{tikz}
\usepackage{forest}
\usepackage{tikz-qtree}
\usepackage{algorithm}
\usepackage{algpseudocode}
\usetikzlibrary{hobby}
\usepackage{pgfplotstable}
\usepackage{tabularx}
\usepackage{booktabs}
\usepackage{multirow}
\usepackage[T1]{fontenc}
\usepackage{pgfplots}
\usepackage{xcolor}
\usepackage{graphicx}
\usepackage{colortbl}
\usepackage{array}
\usepackage{pifont}
\usepackage{adjustbox}
\usepackage{caption}
\usepackage{makecell}
\usepackage{setspace}
\usepackage{enumitem}
\usepackage{authblk}
\usepackage{gensymb}
\def\BibTeX{{\rm B\kern-.05em{\sc i\kern-.025em b}\kern-.08em
    T\kern-.1667em\lower.7ex\hbox{E}\kern-.125emX}}

\pgfplotsset{compat=1.14}
\tikzset{
    rotate around with nodes/.style args={#1:#2}{
        rotate around={#1:#2},
        set node rotation={#1},
    },
    rotate with/.style={rotate=\qrrNodeRotation},
    set node rotation/.store in=\qrrNodeRotation,
}

\tikzset{edge from parent/.style=
{draw, thick, blue, edge from parent path={(\tikzparentnode.south)
-- +(0,-4pt)
-| (\tikzchildnode)}},
blank/.style={draw=none}}

\xdefinecolor{gray95}{gray}{0.65}
\xdefinecolor{gray25}{gray}{0.8}
\newcommand{\cmark}{\ding{51}}%
\newcommand{\xmark}{\ding{55}}%

\begin{document}

%
% paper title
% Titles are generally capitalized except for words such as a, an, and, as,
% at, but, by, for, in, nor, of, on, or, the, to and up, which are usually
% not capitalized unless they are the first or last word of the title.
% Linebreaks \\ can be used within to get better formatting as desired.
% Do not put math or special symbols in the title.
\title{Coverage Protocols for Wireless Sensor Networks:\\ Review and Future Directions}

\author{Riham Elhabyan, Wei Shi and Marc St-Hilaire \thanks{Riham Elhabyan is with the Department of Fisheries and Oceans, Canada. Wei Shi and Marc St-Hilaire are with the School of Information Technology, Faculty of Engineering and Design, Carleton University, email: riham.elhabyan@dfo-mpo.gc.ca, \{wei.shi,marc.sthilaire\}@carleton.ca} \thanks{Riham Elhabyan is  the corresponding author.}}

\markboth{JOURNAL OF COMMUNICATIONS AND NETWORKS, VOL. 17, NO. 4, January 2019}
{Elhabyan \lowercase{\textit{et al}}.: Coverage Protocols for Wireless Sensor Networks: Review and Future Directions}

% The only time the second header will appear is for the odd numbered pages
% after the title page when using the twoside option.
% 
% *** Note that you probably will NOT want to include the author's ***
% *** name in the headers of peer review papers.                   ***
% You can use \ifCLASSOPTIONpeerreview for conditional compilation here if
% you desire.

% If you want to put a publisher's ID mark on the page you can do it like
% this:
%\IEEEpubid{0000--0000/00\$00.00~\copyright~2015 IEEE}
% Remember, if you use this you must call \IEEEpubidadjcol in the second
% column for its text to clear the IEEEpubid mark.

% use for special paper notices
%\IEEEspecialpapernotice{(Invited Paper)}

% make the title area
\maketitle

% As a general rule, do not put math, special symbols or citations
% in the abstract or keywords.
\begin{abstract}
The coverage problem in wireless sensor networks (WSNs) can be generally defined as a measure of how effectively a network field is monitored by its sensor nodes. This problem has attracted a lot of interest over the years and as a result, many coverage protocols were proposed. In this survey, we first propose a taxonomy for classifying coverage protocols in WSNs. Then, we classify the coverage protocols into three categories (i.e. coverage-aware deployment protocols, sleep scheduling protocols for flat networks, and cluster-based sleep scheduling protocols) based on the network stage where the coverage is optimized. For each category, relevant protocols are thoroughly reviewed and classified based on the adopted coverage techniques. Finally, we discuss open issues (and recommend future directions to resolve them) associated with the design of realistic coverage protocols. Issues such as realistic sensing models, realistic energy consumption models, realistic connectivity models and sensor localization are covered.

\end{abstract}

% Note that keywords are not normally used for peerreview papers.

% For peer review papers, you can put extra information on the cover
% page as needed:
% \ifCLASSOPTIONpeerreview
% \begin{center} \bfseries EDICS Category: 3-BBND \end{center}
% \fi
%
% For peerreview papers, this IEEEtran command inserts a page break and
% creates the second title. It will be ignored for other modes.

\section{Introduction}
Wireless sensor networks (WSNs) have attracted significant attention from the research community and industry in the last few years. The main reason for the recent research efforts and rapid development of WSNs is their potential application in a wide range of contexts including military operations, environment monitoring, surveillance systems, health care, and public safety \cite{wsnbk1} \cite{routinginwsn}. These applications require the deployment of a number of sensors to cover a given region of interest (ROI) in the network field. Although sensor nodes can work autonomously, they can also work collaboratively to monitor the physical parameters of an environment. Sensor nodes can sense the environment, communicate with neighboring nodes, and in many cases, perform basic computations on the data being collected \cite{swsurvey} \cite{Akkaya2005325}. These features make WSNs an excellent choice for many applications \cite{routinginwsn} running in environments that are hazardous for human presence.

The coverage problem is one of the fundamental problems in WSNs as it has a direct impact on the sensors energy consumption and the network lifetime \cite{wang}. The coverage problem can generally refer to how to monitor the network field effectively.

There are several ways to classify the coverage problems in WSNs. Coverage problems can be classified, according to the frequency of network field monitor, into either continuous coverage problems or sweep coverage problems. Continuous coverage problems can be further classified, according to the region of interest for monitoring, into three types: area coverage, point coverage, and barrier coverage. Furthermore, coverage problems can be classified, according to the required coverage degree, into either 1-coverage problems or K-coverage problems.

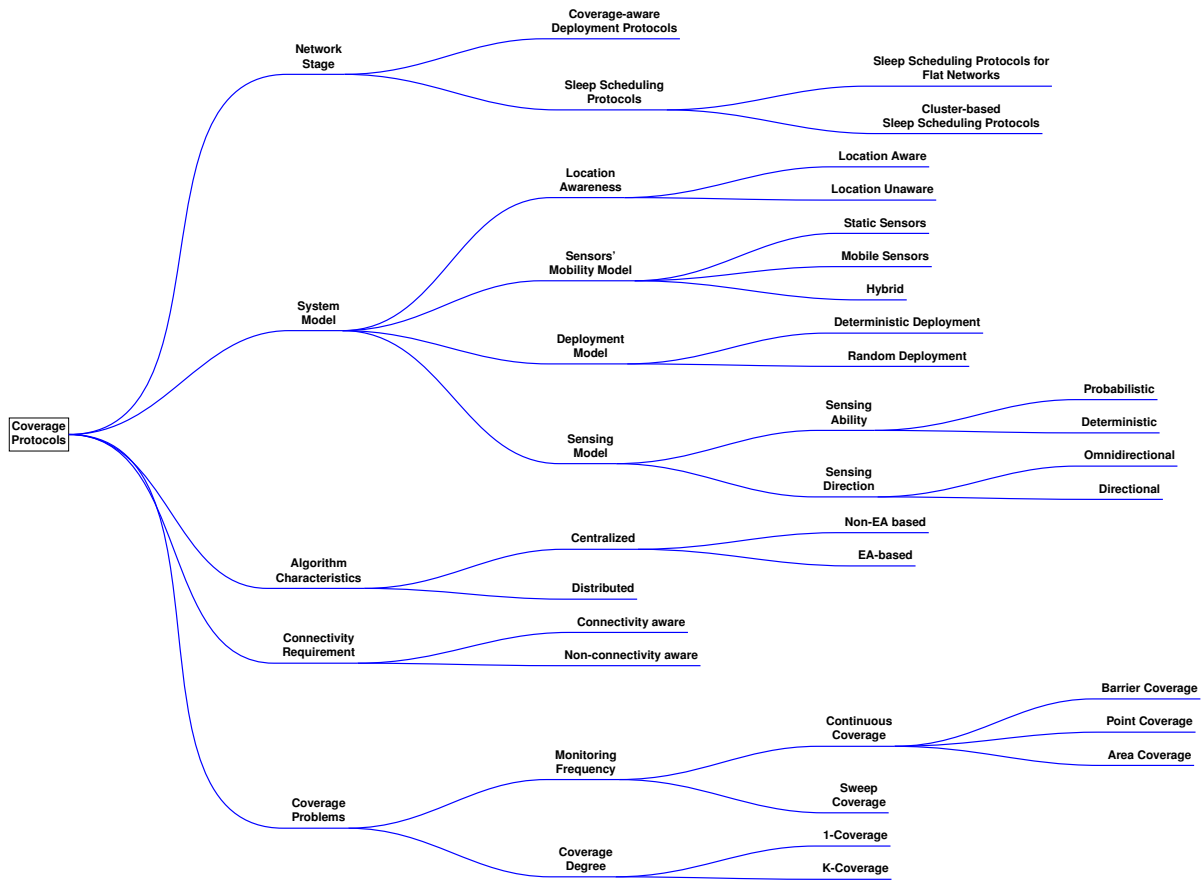
\begin{figure*}
\centering
\scalebox{0.45}{%
 \begin{forest}
    for tree={
     font=\sffamily\bfseries,
      fit=band,
      grow=east,
      parent anchor=south east,
      child anchor=south west,
      align=center,
      inner sep=2,
      l sep+=160pt,
      s sep+=5pt,
      edge={rounded corners=5pt, line width=1pt},
      edge path={
        \noexpand\path [draw, blue,\forestoption{edge}] (!u.parent anchor) [out=0, in=180] to (.child anchor)\forestoption{edge label} -- (.south east);
      },
      for root={
      parent anchor=east,
      rectangle,
      draw,
      },
    }
    [Coverage\\ Protocols
     [Coverage \\Problems
       [Coverage\\Degree
        [K-Coverage
          ]
          [1-Coverage
          ]
        ]
        [Monitoring\\ Frequency
        [Sweep \\Coverage
        ]
          [Continuous \\ Coverage
            [Area Coverage]
            [Point Coverage]
            [Barrier Coverage]
          ]
       ]
      ]
       [Connectivity\\ Requirement
          [Non-connectivity aware 
          ]
          [Connectivity aware
          ]
        ]
      [Algorithm\\Characteristics
        [Distributed
        ]
        [Centralized
          [EA-based
          ]
          [Non-EA based
          ]
        ]
      ]
      [System\\ Model
      [Sensing\\Model 
      [Sensing\\ Direction
          [Directional
          ]
          [Omnidirectional
          ]
        ]
            [Sensing\\ Ability
          [Deterministic
          ]
          [Probabilistic
          ]
        ]
        ]
         [Deployment\\Model 
         [Random Deployment
        ]
      [Deterministic Deployment
        ]
      ]
      [Sensors'\\Mobility Model 
          [Hybrid 
        ]
        [Mobile Sensors
        ]
         [Static Sensors
        ]
       ]
      [Location\\Awareness
          [Location Unaware
          ]
          [Location Aware
          ]
        ]
      ]
      [Network\\ Stage
          [Sleep Scheduling\\ Protocols
          [Cluster-based\\Sleep Scheduling Protocols
          ]
           [Sleep Scheduling Protocols for\\ Flat Networks
          ]
          ]
          [Coverage-aware\\Deployment Protocols
          ]
      ]
     ]
  \end{forest}
  }
   
\caption{Taxonomy for classifying coverage protocols in WSNs}
\label{tax}
\end{figure*}

On the other hand, coverage protocols can be classified based on the connectivity requirement, to either connectivity aware coverage protocols or non-connectivity aware coverage protocols. Furthermore, coverage protocols can be classified, according to the adopted algorithm characteristics, into either distributed protocols or centralized protocols. Centralized coverage protocols can be further classified into either evolutionary algorithm (EA) based protocols or non-EA based protocols. Moreover, coverage protocols can be classified according to the system model of the network. There are four features under the system model: sensor location awareness (aware or unaware), sensor mobility models (static, mobile or hybrid of both), sensor deployment models (deterministic or random), and sensor sensing model. Sensing models are broadly classified, based on the sensing ability, into two types: deterministic sensing models and probabilistic sensing models. Sensing models can also be classified, based on the direction of the sensing range, into either directional sensing models or omnidirectional sensing models. Coverage protocols can also be classified based on when the coverage optimization happens, i.e. into either coverage-aware deployment protocols, when coverage optimization happens before the deployment stage, or sleep scheduling protocols, when coverage optimization happens after the deployment stage. Sleep scheduling protocols can be further classified, based on the network topology, into either cluster-based sleep scheduling protocols or sleep scheduling protocols for flat networks. A detailed description of the various dimensions of the classification discussed above is given in Section~\ref{pre}. Fig.~\ref{tax} shows the taxonomy for classifying coverage protocols in WSNs.

\subsection{Related Reviews}

Several survey papers related to the coverage issue in WSNs exist in the literature. However, to the best of our knowledge, none of the existing studies analyze, review and provide a clear description of all features that cover all factors as well as classify the coverage problems in its entirety. Most of these surveys focus on a subset of features for classifications and overlook the others. In the next paragraphs, we briefly discuss the content of each survey paper and we highlight how this survey is different and more comprehensive than the previous ones.

A specific review of energy-efficient coverage protocols in WSNs is done in \cite{survey2016}. In this review, the authors present the basic knowledge of the coverage problem in WSNs. The reviewed coverage protocols are broadly classified based on the protocol characteristics (distributed vs. clustered) and the sensor nodes location information (location-aware vs. location-unaware). The review considers area coverage protocols only, for both deterministic sensor nodes deployment and random sensor nodes deployment. Finally, the review focuses on coverage protocols that use static nodes only and that are based on the boolean sensing model (a detailed sensing model classification is presented in Section~\ref{SModels}).

The relationship between coverage and connectivity in WSNs is analyzed in \cite{Zhu2012619}. In this review, the coverage protocols are classified into three categories: coverage deployment strategies, sleep scheduling mechanisms, and adjustable coverage radius protocols. This survey however, mainly focuses on coverage protocols that adopt the boolean sensing model.

A comprehensive survey on barrier coverage in WSNs is given in \cite{Wu2016}. The reviewed barrier coverage protocols are mainly classified into two categories: barrier coverage for static sensor nodes and barrier coverage for mobile sensor nodes. The protocols are further classified based on the following criteria: the sensing range direction (omnidirectional vs. directional), the sensing model (boolean, probabilistic and full-view), and the coverage requirement (weak \textit{k}-barrier coverage vs. strong \textit{k}-barrier coverage). Moreover, several optimization problems in barrier coverage are studied.

Another review of barrier coverage is given in \cite{survey2016d}. However, the focus of this review is barrier coverage for directional sensor nodes only. The examined protocols are classified, based on the coverage requirement, into four categories: strong barrier and weak barrier, \textit{1}-barrier and \textit{k}-barrier, worst and best-case coverage and exposure path coverage, and any-view coverage and full-view coverage.

In \cite{tc}, the coverage issue is discussed as a topology control technique in WSNs. The studied coverage protocols are classified into three categories respectively: area coverage protocols, barrier coverage protocols and sweep coverage protocols. Area coverage protocols are further classified based on the types (i.e. static, mobile or hybrid) of sensors available in the WSNs and the coverage requirement (\textit{1}-coverage or \textit{k}-coverage). Moreover, barrier coverage protocols are studied for both deterministic and probabilistic sensing models.

The author of \cite{kc} presents a brief survey on \textit{k}-coverage problems and protocols. The protocols were mainly classified, %based on the \textit{k}-coverage problem type, 
into two categories: \textit{k}-coverage verification protocols and sleep scheduling protocols for \textit{k}-coverage problems.

A review of evolutionary algorithm (EA)-based sleep scheduling protocols is given in \cite{Musilek2015100}. The authors highlight the main reasons behind adopting EA in sleep scheduling protocols. Moreover, the reviewed sleep scheduling protocols are classified, based on the EA they adopted, into four categories: swarm intelligence (ant colony optimization (ACO), particle swarm optimization (PSO), and pulse-coupled biological oscillators (PBO)) protocols, genetic algorithms (GA), differential evolution (DE), cellular automata, and protocols which uses other types of EAs. 

In this paper, we first present a list of features and their values, with which we develop a taxonomy for classifying coverage protocols in WSNs comprehensively. We then review a broad range of coverage protocols conceived in WSNs, mainly categorized into the following three groups: coverage aware deployment protocols, sleep scheduling protocols for flat networks, and cluster-based sleep scheduling protocols. For each group, we compare different coverage protocols on their features identified in Figure \ref{tax}. Table \ref{rsumm} highlights the main differences between the aforementioned review papers and ours in terms of features discussed.

\begingroup
    \renewcommand*{\arraystretch}{1.5}%
    \definecolor{tabred}{RGB}{230,36,0}%
    \definecolor{tabgreen}{RGB}{0,116,21}%
    \definecolor{taborange}{RGB}{255,124,0}%
    \definecolor{tabbrown}{RGB}{171,70,0}%
    \definecolor{tabyellow}{RGB}{255,253,169}%
     \newcommand*{\redtriangle}{\textcolor{tabred}{\ding{115}}}%
    \newcommand*{\greenbullet}{\textcolor{tabred}{\ding{115}}}%
    \newcommand*{\orangecirc}{\textcolor{tabred}{\ding{115}}}%
    \newcommand*{\browncirc}{\textcolor{tabred}{\ding{115}}}%
    \newcommand*{\headformat}[1]{{\small#1}}%
    \newcommand*{\vcorr}{%
      \vadjust{\vspace{-\dp\csname @arstrutbox\endcsname}}%
      \global\let\vcorr\relax
    }%
    \newcommand*{\HeadAux}[1]{%
      \multicolumn{1}{@{}r@{}}{%
        \vcorr
        \sbox0{\headformat{\strut #1}}%
        \sbox2{\headformat{Complex Data Movement}}%
        \sbox4{\kern\tabcolsep\redtriangle\kern\tabcolsep}%
        \sbox6{\rotatebox{90}{\rule{0pt}{\dimexpr\ht0+\dp0\relax}}}%
        \sbox0{\raisebox{\dimexpr\dp0-\ht0\relax}[0pt][0pt]{\unhcopy0}}%
        \kern.75\wd4 %
        \rlap{%
          \raisebox{\wd4}{\rotatebox{90}{\unhcopy0}}%
        }%
        \kern.25\wd4 %
        \ifx\HeadLine Y%
          \dimen0=\dimexpr\wd2+.5\wd4\relax
          \rlap{\rotatebox{90}{\hbox{\vrule width\dimen0 height .4pt}}}%
        \fi
      }%
    }%
    \newcommand*{\head}[1]{\HeadAux{\global\let\HeadLine=Y#1}}%
    \newcommand*{\headNoLine}[1]{\HeadAux{\global\let\HeadLine=N#1}}%
    \noindent
    \setlength{\tabcolsep}{6pt}
     
    \begin{table} [H]
   
  \caption{A comparison between the related reviews and our review, in terms of different dimensions of the surveyed protocols'}
   \resizebox{\columnwidth}{!}{
           \begin{tabular}{%
      >{\bfseries}lc|>{}c|
      *{17}{c|}>{}c
    }%
    &
      \head{\bfseries Year Published} &
       \head{\bfseries{Area Coverage}} &
      \head{\bfseries{Point Coverage}} &
      \head{\bfseries{Barrier Coverage}} &
      \head{\bfseries{Sweep Coverage}} &
      \head{\bfseries{K-coverage}} &
      \head{\bfseries{Sleep Scheduling}} &
       \head{\bfseries{Connectivity}} &
      \head{\bfseries{Cluster-based Coverage}} &
      \head{\bfseries{Sensors Deployment}} &
      \head{\bfseries{Boolean}} &
      \head{\bfseries{Probabilistic}} &
      \head{\bfseries{Directional}} &
      \head{\bfseries{Omnidirectional}} &
      \head{\bfseries{Distributed}} &
      \head{\bfseries{Centralized}}  &
      \head{\bfseries{EA-based}}  
       \\
       
      \sbox0{S}%
      \rule{0pt}{\dimexpr\ht0 + 2ex\relax}%
       \cite{survey2016}  & {\bfseries 2016}   & \redtriangle  &  &   &    &  & \redtriangle  & \redtriangle & \redtriangle  &  & \redtriangle  &  &   & \redtriangle & \redtriangle & & \\\hline
     \cite{Zhu2012619}  & {\bfseries 2012}   & \redtriangle  & \redtriangle &  \redtriangle & \redtriangle   & \redtriangle & \redtriangle  & \redtriangle &  & \redtriangle & \redtriangle  &  &   & \redtriangle & \redtriangle &\redtriangle & \\\hline
     \cite{Wu2016}  & {\bfseries 2016}   &   &  &  \redtriangle &    & \redtriangle & \redtriangle  &  &   & \redtriangle & \redtriangle  & \redtriangle & \redtriangle  & \redtriangle & \redtriangle &\redtriangle & \\\hline
      \cite{survey2016d}  & {\bfseries 2016}   &   &  &  \redtriangle &    & \redtriangle & \redtriangle  &  &   & \redtriangle & \redtriangle  &  & \redtriangle  &  & \redtriangle &\redtriangle & \\\hline
      \cite{tc}   & {\bfseries 2013}   & \redtriangle  & \redtriangle &  \redtriangle & \redtriangle   & \redtriangle & \redtriangle  & \redtriangle &   & \redtriangle & \redtriangle  &  &   & \redtriangle & \redtriangle &\redtriangle & \\\hline
   \cite{kc}  & {\bfseries 2014}   & \redtriangle  &  &   &    & \redtriangle & \redtriangle  & \redtriangle &   & \redtriangle & \redtriangle  &  &   & \redtriangle & \redtriangle &\redtriangle & \\\hline
     \cite{Musilek2015100}   & {\bfseries 2015}   & \redtriangle  & \redtriangle &   &    & \redtriangle & \redtriangle  & \redtriangle & \redtriangle  &  & \redtriangle  &  &   & \redtriangle &  &\redtriangle & \redtriangle\\\hline
       \rowcolor{tabyellow}
     This Review   & {\bfseries 2018}   & \redtriangle  & \redtriangle & \redtriangle  &   \redtriangle & \redtriangle & \redtriangle  & \redtriangle & \redtriangle  & \redtriangle & \redtriangle  & \redtriangle & \redtriangle  & \redtriangle &\redtriangle  &\redtriangle & \redtriangle\\\hline
      \\[.2ex]
     \end{tabular}%
    }
    \label{rsumm}
     
        \end{table}
   \endgroup

\subsection{Our Contributions}
%In this paper, we present a thorough and up to date review of the coverage problem in WSNs. Moreover, we discuss the open issues associated with the design of realistic energy-efficient coverage protocols and the future directions to resolve them.

%Many review papers that examine the coverage problem in WSNs exist in the literature. Without loss of generality, most of the existing reviews concentrated on discussing coverage protocols that are designated to solve a specific coverage problem. Moreover, many of these reviews focused on the boolean sensing model. 

In this paper, we discuss a wide range of coverage problems and coverage protocols. %We also discuss protocols that were proposed recently to solve these problems.
The objective of this review is to provide a better understanding of different classifications of coverage protocols in WSNs and to stimulate new research directions in this area. %Moreover, we aim at motivating the researchers in this field to lower the barrier to designing and developing realistic coverage protocols. In order to achieve that, we discussed and recommended the development of a realistic sensing model which reflect the radio irregularity problem in WSNs. 
More specifically, we discuss and compare coverage protocols in various sensing models, which includes a realistic sensing model that reflects the radio irregularity in WSNs. Our contributions are four-folds:
\begin{itemize}%[noitemsep, nolistsep]
\item We present a broad discussion and a clear classification of various coverage protocols conceived in WSNs.
\item We provide an in-depth review of the up to date protocols designed to solve different coverage problems.
\item We give a thorough discussion on the open issues associated with the design of realistic energy-efficient coverage protocols for WSNs.
\item We recommend potential future directions to solve some unrealistic assumptions that were assumed by many of the previously proposed protocols.
\end{itemize}

\subsection{Paper Organization}

The remainder of this paper is organized as follows. Section~\ref{pre} gives the necessary background knowledge about the coverage problems in WSNs. From Section~\ref{CADP} to Section~\ref{CAClP}, we present related work on coverage protocols based on the network stage where the coverage is optimized. The reviewed protocols are classified into three categories: coverage-aware deployment protocols, sleep scheduling coverage protocols for flat networks, and cluster-based coverage protocols. Section~\ref{OI} provides a detailed discussion of the open research issues to be tackled. The discussion covers the following four directions: 
\begin{itemize}
\item Realistic sensing model
\item Realistic coverage-aware clustering protocols
\item Realistic connectivity model
\item Sensors localization
\end{itemize}  
Finally, Section~\ref{CFD} concludes this paper and highlights future research directions.

\section{Preliminaries}
\label{pre}
In this Section, we give the necessary background knowledge about the coverage issue in WSNs. First, we discuss the different design factors in developing coverage protocols. Then, we present three different types of coverage problems in WSNs. Finally, we outline different sensing models that are used to model the sensor coverage in WSNs.

A list of acronyms used in this paper is displayed in Table \ref{acr}.

\begin{table}[h]
\centering
\caption{List of Acronyms}
\resizebox{\columnwidth}{!}
{
\begin{tabular}{ll}
\toprule[1.5pt]
 \textbf{Acronym} & \textbf{Description}\\
\midrule
   WSNs & Wireless Sensor Networks\\
   ROI & Region Of Interest\\
   POI & Point Of Interest\\
   EA & Evolutionary Algorithm\\
   ACO & Ant Colony Optimization\\
   PSO & Particle Swarm Optimization\\
   PBO & Pulse-coupled Biological Oscillators\\
   GA & Genetic Algorithm\\
   DE & Differential Evolution\\
   CA& Cellular Automata\\
   DE & Differential Evolution\\
   BS& Base Station\\
   CH& Cluster Head\\
   NP & Non Polynomial\\
   OCP & Optimal Coverage Problem\\
   DSC & Disjoint Sets of Covers\\
   PDR & Packet Delivery Rate\\
   LQI & Link Quality Indicator\\
   RSSI & Received Signal Strength Indicator\\
   MCSDP & Maximum Coverage Sensor Deployment Problem\\
   MOEAD& Multi-objective Evolutionary Algorithm (EA)-based on Decomposition
(MOEA/D)\\
  \bottomrule[1.5pt]
\end{tabular}
}
\label{acr}
\end{table}

\subsection{Design Factors}\label{DF}
There are several design factors that have a direct impact on developing coverage protocols for WSNs.

\subsubsection{Coverage Degree}
In its simplest form, coverage means that every point in a ROI is monitored by (i.e. within the sensing range of) at least one sensor. This is referred to as \textit{1-coverage} problem. A more general form is the \textit{$k$-coverage} problem, where each point in a ROI should be monitored by $k$ or more sensors. We also say that the coverage degree of a \textit{$k$-covered} WSNs is $k$. A \textit{$k$-coverage} WSNs is desired in certain applications, such as intrusion detection applications and military applications, because it provides redundancy and therefore enables fault tolerance and stronger monitoring. The coverage degree is considered as one of the main application coverage requirements \cite{wang}.
%##[notes from Wei: what are the application coverage requirements?} I explained that, Done.

\subsubsection{Sensors Deployment Models}
The sensors deployment model is an important design criterion for designing energy-efficient coverage protocols in WSNs. Depending on the application requirements, sensor node deployment can be either random or deterministic \cite{Aznoli2016}.

Random deployment is normally adopted in large-scale applications, where the sensor nodes are scattered randomly into the network field. For instance, sensors can be dropped from an airplane to a remote and hard-to-access area \cite{4753650}. Random deployment is more applicable in dangerous and inaccessible environments.

Deterministic deployment is commonly adopted in small to medium-scale applications, where the number and position of the sensor nodes can be predetermined in advance. The sensors can be deployed either manually or by robots. A regular and symmetric deployment/placement pattern is usually adopted in such deployment methods. Deterministic deployment is suitable only in controlled and human-friendly environments.

\subsubsection{Algorithm Characteristics}
The algorithm adopted by the coverage protocols can be classified into distributed algorithms or centralized algorithms. In distributed algorithms, each sensor node in the network makes a decision about its working mode based on its neighbors' information. Centralized algorithms, on the other hand, require each sensor node to forward its data to a central unit, like a \textit{Base Station (BS)}. Furthermore, most of the coverage problems were defined as NP-hard optimization problems \cite{wang} \cite{7570253}. EAs have been used to solve such problems \cite{Musilek2015100} \cite{Aznoli2016} \cite{Khalesian2016126} \cite{Gupta2015}.

\subsection{Coverage Problems in WSNs}
Coverage problems can be classified, based on the frequency of monitoring the network field, into continuous coverage problems or periodical coverage problems. Moreover, there are other WSNs problems which can affect or get affected by the coverage problem.

\subsubsection{Continuous Coverage Problems}
Continuous coverage problems arise in many WSNs applications that require continuous monitoring of the network field. These problems can be further classified, based on the monitoring requirements, into three types: area coverage, point coverage, and barrier coverage.

Area/blanket coverage problems arise when the whole sensor field needs to be monitored. In other words, every single point of the network field should be within the sensing range of at least one sensor node. On the other hand, point/target coverage problems are related to monitoring a set of targets or points of interests (POI). Different from area coverage and point coverage, barrier coverage is not concerned with monitoring either the entire ROI or any POI. Instead, barrier coverage is to monitor only the borders of a ROI to detect intruders \cite{wang} \cite{Wu2016}.

\subsubsection{Periodical/Sweep Coverage Problems}
There are typical applications where only periodic monitoring is sufficient for a certain set of POIs. \textit{Sweep Coverage} is introduced for such applications. In sweep coverage, the entire field may not be covered all the time. A number of mobile sensor nodes moving within the network field are used to collect data about the POIs and deliver it to a central processing unit, like a BS. Using a combination of static and mobile sensors in sweep coverage is more effective. When static sensor nodes detect these POIs, they record the data locally so that it can be later retrieved by mobile sensor nodes within a delay bound \cite{5361809}. The main goal of the sweep coverage problem is to minimize the number of mobile sensor nodes used and guarantee sweep coverage for a given set of POIs \cite{Gorain20142699} \cite{6021963}.

\subsubsection{Coverage-related Problems}
\label{cvrgrltd}
The connectivity problem is concerned with finding direct or indirect high-quality and energy-efficient communication links between the sensors to a BS. Such list of links would provide efficient and reliable transmission of data. However, coverage solutions can not guarantee network connectivity. Therefore, both the coverage and the connectivity issues should be jointly investigated to ensure proper deployment.\cite{Zhao20082205} \cite{Goel:2014} \cite{Ghosh2008303}. The deployment is another important problem in WSNs that deals with finding the optimal sensors placement pattern that fulfills both coverage and connectivity requirements.

Cluster-based routing provides an efficient approach to reduce the energy consumption of sensor nodes and maximize the lifetime and scalability of WSNs. Finding the optimal clusters while considering the coverage optimization of the network may lead to more energy-efficient solutions for WSNs. \cite{Faheem2015309} \cite{Gu2014384}.

\subsection{Sensing Models}
\label{SModels}
In WSNs, each sensor node has a limited sensing range, and hence can only cover a limited physical area of the network field. Sensing models are abstraction models that are used to reflect the sensors' sensing ability and quality \cite{Wang2010}. The sensing models can be classified, based on the direction of the sensing range, into either directional sensing models or omnidirectional sensing models. Moreover, sensing models are broadly classified, based on the sensing ability, into two types: deterministic sensing models and probabilistic sensing models. 
\subsubsection{Directional Sensing Model}
In directional sensing model \cite{Ma:2007} \cite{Ma2005}, each sensor has a finite angle of view and cannot sense the whole circular area around it. Directional sensor nodes may have several working directions and may adjust their sensing directions during their operation. In this model, the sensing area, also known as the field of view, is a sector described by four parameters $L, R, V, \alpha$ where $L$ is the sensor node position, $R$ is the sensing radius, $V$ is the working direction and $\alpha$ is the angle of view. A point $Q$ is said to be covered by a directional sensor $S$ if and only if the following conditions are met \cite{Ma:2007}
\begin{itemize}[noitemsep,nolistsep]
\item $d(L,Q) \leq R$ where $d(L,Q)$ is the Euclidean distance between the points $L$ and $Q$ and
\item The horizontal angle of $ \overrightarrow{LQ}$ is within $[-\alpha, \alpha ]$
\end{itemize}

Infrared, ultrasound, and video sensors are such examples of directional sensors.

%### [NOTES FROM WEI: This reference must be added: http://link.springer.com/chapter/10.1007/11599463_70   and formal definition of directional sensing model must be used] Done

\subsubsection{Omnidirectional Sensing Model}
Omnidirectional sensing model is a special case of directional sensing model \cite{Ma2005} %### [NOTES FROM WEI: quot the same above-mentioned paper ] Done
 where $\alpha = 360\degree$. Omnidirectional sensors cover a unit of a circle and they have only one working direction. Many legacy sensor nodes equipped with temperature, humidity, and magnetic sensors are omnidirectional sensors.

\subsubsection{Boolean Sensing Model}
Boolean (deterministic/disk) sensing model is the simplest and most commonly used sensing model \cite{6867778}. In this model, if a point (or event) $P$ in the network field is located within the sensing range $R$ of sensor node $S$, then it is assumed that $P$ is covered/detected by $S$. The sensing area of $S$ is defined as a disk centered at $S$ with a radius of the sensing range $R$.

In this model, the coverage function, $C(S,P)$, of sensor node $S$ and point $P$ is given by the following equation:
\begin{equation}
C(S,P) = \begin{cases}
1,& \text{if }d(S,P) \leq R\\
0,& \text{otherwise}
\end{cases}
\label{sense1eq}
\end{equation}

Where $d(S,P)$ is the Euclidean distance between sensor node $S$ and point $P$.

\subsubsection{Probabilistic Sensing Models}
The probabilistic sensing model was firstly proposed by \cite{Zou} as a more realistic extension of the boolean sensing model. This model was motivated by the fact that sensor detections are usually imprecise and the quality of sensing gradually decreases with increasing distance away from the sensor \cite{Megerian}. Therefore, the coverage function, $C(S, P)$, needs to be expressed in probabilistic terms. The probabilistic sensing model is further classified into two models: the Elfes sensing model and the shadow fading sensing model.

\paragraph{\textbf{The Elfes Sensing Model}}
\label{elfes}

%######[notes from Wei: this paper must be cited: A. Elfes, ?Occupancy grids: a stochastic spatial representation for active robot perception,? in Autonomous Mobile Robots: Perception, Mapping and Navigation, vol. 1, S. S. Iyenger and A. Elfes, Editors, IEEE Computer Society Press, 1991, pp. 60-70.] Done

Two sensing radii are defined in the Elfes sensing model \cite{30720}, $R_{min}$ and $R_{max}$ where $R_{min}$ defines the starting of the uncertainty in the sensor detection. If a point (or event) $P$ in the network field is located within the sensing range $R_{min}$ of sensor node $S$, then it is assumed that $P$ is definitely covered/detected by $S$. If point $P$ is located beyond the sensing range $R_{max}$, then it is definitely not covered/detected by $S$. Otherwise, point $P$ is covered/detected by $S$ with probability $p$. The coverage function, $C(S, P)$, is given in the following equation:
\begin{equation}
\resizebox{.9\hsize}{!}
{$
C(S, P) = \begin{cases}
1,& \text{if }d(S, P) \leq R_{min}\\
p = e^{-\lambda (d(S, P)- R_{min})^\gamma},& \text{if }{R_{min} < d(S, P) < R_{max}}\\
0,& \text{if } d(S,P) \geq R_{max}
\end{cases}
$}
\label{sense2eq}
\end{equation}

Where the $\lambda$ and $\gamma$ parameters are adjusted according to the physical properties of the sensor. It should be noted that the Elfes sensing model is considered a more general model, where it becomes a boolean sensing model when $R_{min}=R_{max}$.

\paragraph{\textbf{The Shadow Fading Sensing Model}} In the aforementioned sensing models, the sensing radius of a sensor node has a constant value in all directions around it. Therefore, its sensing ability depends only on the distance between the sensor node and the point of interest. However, obstructions in the network field, such as buildings, railway track, power stations and mines, result in extra loss and more variation in the received signal power of both the sensing signal and the signal emitted from targets or events. Moreover, the sensing ability of a sensor is non-uniform and asymmetric in all directions around the sensor due to their hardware configuration and software implementation. Therefore, the sensing radius of a sensor node should not be modelled uniformly in all directions since signals from different directions, corresponding to different propagation paths, suffer from different amounts of \textit{shadowing} loss. The variations in the received signal strength due to obstructions in propagation path is known as \textit{shadowing}.

The shadow fading sensing model given in \cite{tsai} is the first model to consider the impact of the shadowing effects on the coverage problem in WSNs. In this model, the coverage function, $C(S, P)$, represents the probability that sensor $S$ covers/detects $P$ in a shadowed environment. This coverage probability depends on the shadowing fading parameter $\sigma$, the Euclidean distance $d(S, P)$ between sensor node $S$ and point $P$ as well as an average sensing radius $\bar{R}$. The coverage function, $C(S, P)$, is given by the following equation \cite{tsai} \cite{ses}:

\begin{equation}
C(S, P) = \dfrac{1}{A}\int_0^{R_{max}} Q(\dfrac{10 \beta \log_{10}(\dfrac{d(S, P)}{\bar{R}})}{\sigma}) \times 2 \pi d(S, P) dr
\label{sense3eq}
\end{equation}

\begin{figure}[H]
%\centering
\captionsetup{justification=centering}
\subfigure[Deterministic \newline sensing model]{      
       \includegraphics[scale=.8]{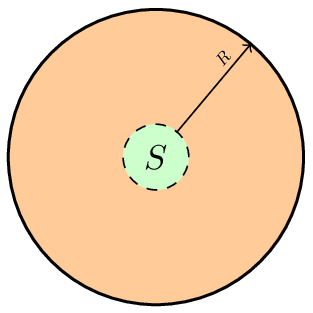}
      \label{sense1}
       }
       \subfigure[Elfes sensing model]{      
       \includegraphics[scale=.8]{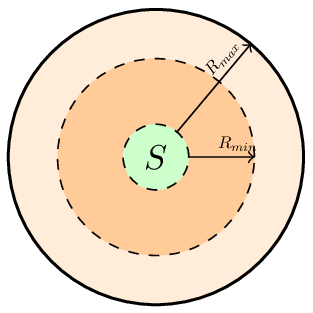}
       \label{sense2}
       }
       \subfigure[Shadow fading \newline sensing model]{      
       \includegraphics[scale=.8]{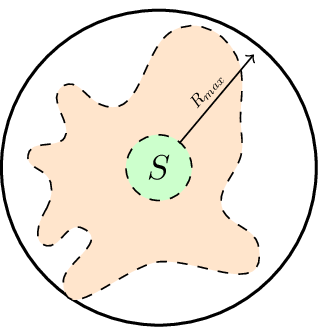}
       \label{shadow}
       }
     \caption{The shape of the sensing area for \newline different sensing models}
     \label{area}
\end{figure}

Where $\beta$ denotes the signal power decay factor and $R_{max}$ denotes the maximum practicable sensing range. $dr$ %###[notes from Wei: dr need to be better explained. ] Done
is a small increment in distance $d(S, P)$ which represents a small differenece in the distance due to the sensor size. More details about these variables or how to calculate $\bar{R}$ can be found in \cite{tsai} \cite{ses}. Fig. \ref{area} illustrates the shape of the sensing area for the aforementioned sensing models.

%\section{Literature Review}
%======================================================================
%Coverage techniques have been studied extensively to improve the performance of WSNs \cite{survey2016, survey2012, tc}. 

In this survey, we classify coverage protocols, based on the network stage where the coverage is optimized, into either coverage aware deployment protocols or sleep-scheduling protocols. Sleep-scheduling protocols are further classified, based on the network topology, into either cluster-based sleep scheduling protocols or sleep scheduling protocols for flat networks, as shown in Fig. \ref{tax}.  The following three sections give a detailed review of such protocols.

\section{Coverage-aware Deployment Protocols}
\label{CADP}
Coverage-aware optimal sensors deployment can be defined as the process of determining the optimal locations of sensors in a network field such that the coverage requirement of an application is met. The coverage hole problem, which refers to finding regions that are not covered by any sensor, is a sub-problem of deployment protocols \cite{Tsai20151305}. Mobile sensors are used to solve such problem by adapting their position in order to fill up sensing holes and eventually increase the area coverage \cite{5724917}. The Maximum Coverage Sensor Deployment Problem (MCSDP) is an example of deployment problems that aims at finding the minimum number of sensors to achieve maximum coverage of the surveillance area. Most deployment problems are NP-hard problems with many conflicting objectives. Therefore, centralized evolutionary approaches are often used to solve various deployment problems \cite{Aznoli2016} \cite{Khalesian2016126} \cite{Gupta2015}.

A PSO-based deployment algorithm, PSODA, is proposed in \cite{Senouci} to solve the deterministic deployment problem for point coverage in WSNs. In PSODA, the MCSDP is modelled as a constrained optimization problem and the main objective of the algorithm is to minimize the number of sensors while satisfying the coverage constraints for all the target points. The ROI is divided into small cells and the center of each cell is a potential position for a sensor. PSODA contains one binary $0/1$ decision variable for each position in the network area where the value of $1$ indicates that a sensor should be deployed at this position and $0$ indicates the opposite. The fitness function uses a weighted-sum approach that combines two sub-objectives: the first one is used to minimize the number of sensors to be deployed and the second one is used to minimize the dissatisfaction of the coverage constraints.%###[notes from Wei April 19th: why the number of sensors is in the weighted-sum?, I rewote it, the goal is to find the min number of sensors to be deployed such thet they provide full coverage, Done.
PSODA assumes that the sensors follow the Elfes sensing model and all the sensors are static and homogeneous. A modified PSO which uses a new position updating procedure for a faster convergence was adopted to solve the premature convergence problem of traditional PSO. Although PSODA was primarily developed to solve the point coverage problem, it can be adopted for applications that require full area coverage. It should be noted that the PSODA protocol does not consider the connectivity between the sensors and the BS.

A constrained Pareto-based Multi-objective Evolutionary Approach, CPMEA, is proposed in \cite{Khalesian2016126} to solve the deterministic deployment problem in WSNs. Unlike PSODA, CPMEA treats the coverage requirement as an objective rather than a constraint. Moreover, CPMEA aims at maintaining the full connectivity between each sensor node and the BS by modelling the connectivity requirement as a constraint. CPMEA uses the Pareto-dominance concept to formulate the objective functions. The main objective is to find more than one Pareto-optimal sensor-layouts that can maximize the coverage and lifetime simultaneously while maintaining full connectivity between the sensors. The decision variables in CPMEA represent the desired positions of the sensor nodes. However, instead of generating a collection of random layouts without considering the connectivity, the initial population is generated in two steps. The first step consists of generating a number of random tree topologies that connect the BS to the sensor nodes. Then in the second step, the positions of sensor nodes are randomly generated based on the BS %###[notes from Wei: I replaced sink with BS.  Please check whether this is correct.] Yes it is OK. Done.
position and the tree structure. CPMEA assumes that the sensors follow the boolean sensing model and all sensors are static and homogeneous.

A GA-based deployment protocol was proposed in \cite{Gupta2015} to ensure both coverage and connectivity of a given set of targets. The goal of the protocol is to select the minimum number of the potential positions for the sensors such that two requirements are met: $k$-coverage and $m$-connectivity. The objective function was defined as a weighted-sum of three scaled sub-objectives: minimizing the number of deployed sensor nodes, maximizing the total achieved coverage and maximizing the connectivity. Each individual in the GA population has a length equals to the number of the potential positions of the sensors. Each gene can have a value of either $1$ or $0$ to indicate whether a sensor should be installed at that location or not. \cite{Gupta2015} assumes that the sensing range is equal to the communication range and all sensors follow the boolean sensing model. It is also assumed that all the sensors are static and homogeneous.

Another approach for solving the sensors deployment problem in WSNs is to find an optimal deployment pattern. %###[notes from Wei April 19th: what is a "deployment pattern" need to be explained further, Explained in the next sentence, Done.
In this approach, it is assumed that the ROI is divided into virtual grids and every sensor is deployed at the intersection points of the grid. The grid shape can either be square, triangle, hexagon, etc. %###[notes from Wei April 19th: do you mean ".... divided into small regions such as mesh"? Yes, Done.
The goal of the deployment protocol is to estimate the pattern (grid shape) and the optimal distance between the sensors. For example, the authors in \cite{7217790} developed a protocol to address the problem of finding a regular node deployment pattern that uses the minimum number of sensors to provide \textit{k}-coverage and \textit{m}-connectivity. The main idea of the proposed protocol is to find a deployment pattern that satisfies three conditions: the network area is \textit{k}-covered, the sensor nodes are \textit{m}-connected, and the number of deployed sensors is minimized. The main goal of this protocol is to estimate the locations and the optimal distance between sensors for three different deployment patterns: triangle, square, and hexagon. The protocol then chooses the deployment pattern to be used to deploy minimum number of sensors while meeting the coverage and connectivity requirements. The protocol assumes a boolean sensing model and all sensors are static and homogeneous.

Another approach for coverage-aware deployment is to deploy and reposition mobile sensors to meet the coverage requirement of a certain application. MobiBar \cite{Silvestri2017111} is a protocol of such design that is proposed for barrier coverage applications. MobiBar is a distributed deployment protocol that utilizes mobile sensors to construct \textit{k} distinct complete barriers and hence provides \textit{k}-barrier coverage. The goal of MobiBar protocol is to achieve a final deployment that provides the maximum achievable barrier coverage by repositioning the mobile sensors. The authors of MobiBar defined a \textit{baseline} as the line that is parallel to the border of the network area to which other barriers should be constructed parallel to it. MobiBar assumes that sensors located on adjacent barriers are able to communicate. These connected barriers are referred to as the connected barrier component. Each barrier in MobiBar has a priority which decreases as the distance between the baseline and the barrier increases. Initially, all sensors move towards the baseline to increase the connectivity of the network. The first sensor to reach the baseline elects itself as a leader of the connected barrier component. Then this leader chooses at most four neighbor sensors, each of which move to an adjacent barrier position. Precedence is given to barriers with higher priorities. In turn, each of these newly relocated sensors repeat this process to at most four of its neighbor sensors. If a relocated node couldn't find enough neighbor sensors to ask to relocate, it keeps performing a limited multi-hop search till it finds one or more. A relocated sensor may reposition itself only to a barrier with a higher priority if it detects one. Newer barriers are then merged into older barriers by repositioning the sensors in the new barrier to the vacant positions of the older barrier. At the end, a single connected barrier is constructed. All sensor nodes in MobiBar are mobile and it also assumes a boolean sensing model for sensing as well as a perfect disk model for communication.

A Mobile Sink (MS) based Coverage Optimization and Link-stability Estimation Routing (MSCOLER) protocol was proposed in \cite{DAHIYA2018191} to i) restore the coverage and ii) prevent transmission faults in the network. MSCOLER operates in two phases. In the first phase, MSCOLER uses a Grid-based Firefly Simulated Annealing (GFSA) to move the mobile sensors near the coverage holes. In order to do so, the network area is divided into grids and each cell in this grid needs to be monitored by at least 1 sensor. The coverage problem is modeled as a nonconstrained optimization problem with the goal of maximizing the coverage ratio. Firefly Simulated Annealing (FSA) is then used to solve this problem by finding the optimal locations of the mobile sensors to restore the coverage holes. In the second phase of MSCOLER, a Link Stability Estimation Routing (LSER) algorithm is used to find the optimal relay sensors to forward the data to the BS. The optimal links are found by minimizing three link quality indicators: the Expected Transmission Time (ETX),  the Received Signal Strength Indicator (RSSI ), and the Link Quality Indicator (LQI). The authors assume a binary sensing model and the first order energy consumption model. All sensors are homogenous, mobile and location-aware.

%A distributed coverage protocol was proposed in \cite{KHEDR201861} where two distributed algorithms were proposed for coverage hole detection and coverage hole repair respectively while ensuring connectivity. The proposed protocol is divided into three phases: initialization, hole detection, and hole recovery. The initialization phase consists of three steps.  In the first step, each sensor finds the sensing neighbours based on the distance between them. In the second step, each sensor use the algorithm in \cite{Librino:2014} to find the redundant sensors within its vicinity. In the final step, each sensor determines the intersection points with the hole by computing the intersection points with every sensor neighbour. 

Table \ref{rltdd} shows a comparison between the coverage aware deployment protocols discussed above.
\begin{table*}[htp]
\centering
\caption{Comparison of coverage aware deployment protocols}
\resizebox{0.8\textwidth}{!}
{
\begin{tabular}{llllllllllll}
\toprule[1.5pt]
 & \textbf{Coverage}& \textbf{Year} &\textbf{Main} & \textbf{Sensing} &\textbf{Location} & &  \multicolumn{4}{l}{\textbf{Protocol Characteristics}}\\
\cmidrule(l){8-11}
\textbf{} & \textbf{Protocol} & \textbf{Published} & \textbf{Goal(s)} & \textbf{Model} & \textbf{Awareness} &  & \textbf{Dist. } & \textbf{Cent.} & \textbf{EC.}\\
\midrule
 &\cite{Gupta2015} & 2015 & Provide full area $K$-coverage/$M$-connectivity &   Boolean &  N/A& & \xmark &  \cmark &  \cmark \\
   &  \cite{7217790} & 2015 & Provide full area $K$-coverage/$M$-connectivity & Boolean &  N/A & &\xmark & \cmark & \xmark \\
  &   MobiBar \cite{Silvestri2017111} & 2017 & Provide $K$-barrier coverage &   Boolean &   Yes &  & \cmark &  \xmark &  \xmark \\
    &    PSODA \cite{Senouci} &   2016 & Solve the MCSDP/Point coverage &   Elfes &   N/A &   & \xmark &  \cmark &  \cmark \\
    &  CPMEA \cite{Khalesian2016126} & 2016 & Provide full area coverage/connectivity & Boolean  &  N/A & &\xmark & \cmark & \cmark \\
    &  MSCOLER \cite{DAHIYA2018191} & 2018 & Provide targeta K-coverage/connectivity & Boolean  &  Yes & &\xmark & \cmark & \cmark \\
  \bottomrule[1.5pt]
\end{tabular}
}
\label{rltdd}
\end{table*}

\section{Sleep Scheduling Protocols for\\  Flat Networks}
\label{SSCP}

In WSNs, sensors can be deployed randomly and in high density to ensures higher coverage. The random deployment of sensors may result in several close-located (redundant) sensors covering the same area and therefore causing unnecessary energy consumption. Activating only the necessary sensors at any particular moment can save energy. The optimal coverage problem (OCP) in WSNs is defined as finding the fewest number of sensors to monitor a given area while maintaining the coverage ratio requirement of the application. The main approach to solving such problem is to employ sleep scheduling protocols, in which redundant sensors are scheduled to be asleep/deactivated  alternately to minimize energy consumption, and hence increase the overall network lifetime while meeting the coverage requirement.

A distributed and localized sleep scheduling protocol called coverage maximization with sleep scheduling (CMSS) protocol is proposed in \cite{7207001}. CMSS assumes a grid-based deployment with sensor nodes deployed at random in grid cells. Moreover, the sensors are location-aware and homogeneous in terms of their sensing range and communication range. The main goal of CMSS is to minimize the number of active sensors while ensuring the ROI is fully covered. This is achieved by minimizing the number of redundant sensors that monitor the same cell. CMSS assumes a boolean sensing model, where each cell is considered covered by a sensor if all points in that cell are within a sensor's sensing radius. Each sensor has two tables: a) a neighbor table, which records the IDs of its neighbors and b) a covered-cells table, which records the covered cells and associated sensors covering each cell. CMSS operates in rounds. When first round starts, each sensor broadcast its location information and sensors that receive this information adjust their neighbor table accordingly. In subsequent rounds, the covered-cells table is updated at the beginning of each round. Each sensor in the network makes its decision on whether it should stay active or go to sleep mode by applying a back-off timer technique and checking its covered-cells table. %The basic idea of CMSS is to deactivate sensors which do not reduce the coverage area when it enters sleep mode.

Coverage-aware scheduling for optimal placement of sensor \cite{Jamali2015}, (CAOP), is another distributed sleep scheduling protocol. Similar to CMSS, all sensors in CAOP are location-aware and homogeneous in terms of their sensing range. These sensor nodes are randomly deployed. Moreover, sensors are aware of the number of nodes deployed in the network field as well as the network dimensions. %### [notes from Wei: this is what I have added. Please make sure it is indeed the assumption in the paper.] The sensors only know the network size and network dimensions, Done
However, CAOP applies the Elfes sensing model. The main idea of CAOP is that, in the case of deterministic sensor deployment, Polygon is one of the basic placement patterns, in which, sensor nodes are placed at the vertices of polygons. If all vertices embedded within the sensor field are covered, then the whole sensor field is said to be fully covered. However, since sensor nodes are randomly deployed in CAOP, each sensor node makes its own scheduling decision based on the distance between itself and the vertices of the closest polygons. The CAOP protocol operates in rounds. At the beginning of the first round (the decision round), each sensor starts running CAOP independent of other sensors to determine its activity round, i.e. the round at which it will be active. In order to do that, every sensor node in CAOP starts by computing the minimum number of nodes required for covering the whole ROI based on its sensing range and the dimension of the ROI. The assumed places for these nodes are considered as optimal places. Then, each sensor node decides its state (active or sleep) based on the sensor nodes density and its distance from the nearest optimal place. In the subsequent rounds, each sensor check if its activity round matches the current round. At any given round in CAOP, only a subset of the sensors stays in active mode while the other sensors stay in sleep mode waiting for their own round to be active. Each sensor determines its own round number based on its location and the sensing range. Since this process is distributed and a node does not know the location of other nodes except its direct neighbours, it is possible that two nodes that are supposed to cover the same POI decide to go to sleep mode simultaneously. This may consequently create a coverage hole. %### [notes from Wei: I did not read the algorithm, so you need to answer this: is there any chance that two sensors come up with a plan that conflict with each other? E.g. A expects B to sleep/awake and at exactly the same time, B expects A to sleep/awake. Ended up both are out of battery or leaving a hole while both are sleep. If the answer is yes, then we need to point this out as a "concern". If indeed, as i have edited that each sensor nodes know the entire information of the ROI, theoretically each of these nodes will come up with exactly the same result. Please read the algorithm again.] I rewrote the whole paragraph and I explained why is this less likely to happen, Done.

Another common approach that sleep-scheduling protocols adopt is to divide the sensors into disjoint sets of covers (DSCs) such that every set completely covers the entire ROI or a set of targets (e.g. POI) with known locations. Once an active set run out of energy, another set is activated to continue providing the required level of coverage. Network lifetime prolongs when the number of such sets increases. Hence, the goal of this approach is to determine the maximum number of DSCs. Since both the OCP and the DSC problems are well-known NP-hard optimization problems, EAs can be used to solve them \cite{Mostafaei2015} \cite{Abdulhalim2015}.

A variant of PSO, Binary PSO (BPSO), is adopted in a centralized Binary PSO-based sleep scheduling protocol \cite{Zhan2015263} to solve the OCP. BPSO assumes that the sensors are homogeneous, randomly deployed in the network field and adopt the boolean sensing model. The coverage problem was modelled as a constrained $0/1$ programming problem to determine whether a sensor should be in active mode (with value $1$) or in sleep mode (with value $0$). The goal of the protocol is to minimize the number of active sensor nodes while maintaining full area coverage constraint. Moreover, the protocol was extended to find the maximum number of DSCs. This is done by initially minimizing the number of active sensor nodes. These active nodes form the first set and are marked as unavailable. Then, the unassigned sensor nodes form another network topology. This process continues till the last network topology cannot provide full coverage for the area.

A multi-layer GA, (mlGA), is adopted in \cite{Abdulhalim2015} to find the maximum number of DSCs. %###[notes from Wei: I do not understand this following sentence. Do you mean to select the minimum number of sensors to get a maximum number of set covers or do you mean to select the minimum number of sensors and assign this set of sensors to as many set covers as possible?] I rewrote the sentence again, Done
The goal of the \textit{mlGA} protocol is to find the maximum number of DSCs and to ensure that each DSC is assigned the minimum number of sensors which provides full coverage. The mlGA protocol employs a post-heuristic operator, in which the unassigned sensors may be used to enhance the coverage of each set of covering sensors or \textit{set cover} for brevity. Similar to the BPSO protocol, the mlGA protocol identifies the maximum number of DSCs gradually. However, the  mlGA protocol assumes that the sensors adopt the Elfes sensing model to reflect the uncertainty in sensor's sensing ability. It should be noted that the random initialization and update of the population individuals in both \textit{BPSO} and \textit{mlGA} may result in infeasible set cover solutions that do not meet the required coverage constraint. In this case, a repair function is adopted to repair these individuals (i.e. set covers) and hence further move toward the optimal solutions space. The repair function usually works by drawing a random sensor from the set of the unassigned sensors and adding this sensor into an infeasible individual solution. This process continues till the coverage constraint of each infeasible individual is met.

An Energy Efficient Connected Coverage (EECC) algorithm was proposed in \cite{ROSELIN20171} to find the maximal number of non DSCs that ensure target coverage and connectivity while minimizing sensors redundancy around the targets. Authors of EECC argued that non-disjoint cover sets provide a longer network lifetime compared to the disjoint cover set as they may generate more cover sets, which in turn will prolong the network life time. The sensors  are classified into sensing and relay nodes according to its coverage. If the sensor does not cover any of the targets, its coverage is null and it is termed as relay node. A sensor which covers a target is termed as a sensing sensor. Sensing sensors are further classified, based on the number of targets it covers, into three types: single coverage sensor, multi coverage sensor, and critical coverage sensor. Each sensor has a heuristic value derived from its sensing coverage and connectivity to the BS. The value of the coverage heuristic prioritize the sensing node according to its contribution towards coverage while the value of connectivity heuristic prioritize the relay sensor according to its connectivity to sensing nodes and sink. Although the authors have proved that their problem formulation is NP-complete, they have used greedy approach to select the sensors to include in the cover, based on their heauristic values.

A GA-based protocol to find the maximum number of non DSCs to provide K-coverage for a predetermined number of targets was proposed in \cite{ELHOSENY2018142}. Each sensor cover is enough to cover all the targets in the field. The network lifetime of the sensor covers is calculated as the minimum lifetime of a sensor that belongs to that cover, i.e. the sensor that has the minimum remaining energy. The energy-efficient target coverage problem is then formulated as a maximization problem that aims to maximize the aggregated network lifetime among all the sensor covers. Firstly, the adopted GA algorithm determines the optimal cover heads that are responsible for transferring the data to the BS. Then, the algorithm forms the covers based on the coverage range of each sensor, the expected consumed energy, the distance to the BS, and targets positions. Authors assumed that the sensors are mobile and can move freely in the network field , to collect environmental data,  without adhering to any specific sensor mobility model. Moreover, a  cover management method that switches between different sensor covers was proposed. It was also assumed that all the sensors can transmit directly to the BS.

Table \ref{rltdss} shows a comparison between the coverage aware deployment protocols discussed above.
\begin{table*}[htp]
\centering
\caption{Comparison of sleep scheduling protocols for flat networks}
\resizebox{0.8\textwidth}{!}
{
\begin{tabular}{llllllllllll}
\toprule[1.5pt]
 & \textbf{Coverage}& \textbf{Year} &\textbf{Main} & \textbf{Sensing} &\textbf{Location} & &  \multicolumn{4}{l}{\textbf{Protocol Characteristics}}\\
\cmidrule(l){8-11}
\textbf{} & \textbf{Protocol} & \textbf{Published} & \textbf{Goal(s)} & \textbf{Model} & \textbf{Awareness} &  & \textbf{Dist. } & \textbf{Cent.} & \textbf{EC.}\\
\midrule
   & CMSS \cite{7207001} & 2015 &  Minimize \# of active sensors/Full area coverage & Boolean & Yes &  &\cmark & \xmark & \xmark \\
  & CAOP \cite{Jamali2015} & 2015 &   Minimize \# of active sensors/Full area coverage  &   Boolean &  Yes &   & \cmark &  \xmark &  \xmark \\
  & BPSO \cite{Zhan2015263}& 2015 &  Find maximum \# of DSCs/Full area coverage & Boolean & Yes & &\xmark & \cmark & \cmark \\
  & mlGA \cite{Abdulhalim2015} & 2015 &  Find maximum \# of DSCs/Full area coverage &   Elfes &  Yes & & \xmark &  \cmark &  \cmark \\
   & EECC \cite{ROSELIN20171} & 2017&  Find maximum \# of non DSCs/Target coverage &   Boolean &  Yes & & \xmark &  \cmark &  \xmark \\
    & \cite{ELHOSENY2018142} & 2018 &  Find maximum \# of non DSCs/Target coverage &   Boolean &  Yes & & \xmark &  \cmark &  \cmark \\
\bottomrule[1.5pt]
\end{tabular}
}
\label{rltdss}
\end{table*}
\section{ Cluster-based Sleep Scheduling Protocols}
\label{CAClP}
Though both the \textit{Cluster Heads (CHs)} selection problem and the coverage problem have been extensively studied separately, only a few protocols considered them together. Most of existing clustering protocols focus only on selecting CHs to reduce or balance the network's energy consumption, while how to cover the network area effectively is out of the scope of these existing solutions. The following paragraphs describe papers that consider clustering and coverage simultaneously.

A Distributed, Cluster-based Coverage-aware protocol, ECDC, that can be adapted for different applications is proposed in \cite{Gu2014384}. The network area in ECDC consists of randomly deployed static sensor nodes. The main idea behind the protocol is that sensors having higher remaining energy and/or deployed in a densely populated area, and/or cover more POIs are more likely to be selected as CH candidates. Two coverage importance  metrics are introduced to measure the coverage importance for each sensor node: one for the point coverage problem and the other for the area coverage problem. In the point coverage problem, the point coverage importance of a sensor node is determined by the number of POIs covered by that sensor node only. The higher the number of POIs covered by a sensor node, the larger the point coverage importance of that sensor node. In the area coverage problem, the area coverage importance of a node is determined by the number of neighbors. The fewer nodes around a sensor node, the greater the area coverage importance of that sensor node. The clustering process of ECDC is divided into rounds, each of which consists of a cluster set-up phase and a data transmission phase. In the cluster set-up phase, the sensor nodes compete for the CH role based on their relative residual energy and their coverage importance. At the end of this phase, sensor nodes with relatively higher residual energy and smaller coverage importance will be chosen as CHs. In the data transmission phase, a routing tree is constructed to connect the elected CHs to the BS. The CHs aggregate data from their cluster members and then send data to the next hop nodes on the constructed routing tree. It is assumed that the selected CHs are within the communication range of each other and each CH can either send its data directly to the BS or can send its data to a neighboring CH. The ECDC protocol uses a Time Division Multiple Access (TDMA) mechanism to avoid inter-cluster and intra-cluster collisions.

Balanced clustering algorithm (BCA) is another distributed clustering protocol that was proposed in \cite{Shin2015}. Similar to the ECDC protocol, the BCA protocol operates in rounds and favors sensors that are deployed in a densely populated area to act as CHs candidates. Moreover, the BCA protocol creates a set of equally balanced, in terms of their coverages, clusters (i.e. to make the coverage area of each cluster approximately the same). The coverage area of a cluster is defined as the union of the coverage areas of all cluster members. In BCA, each sensor calculates its probability of becoming a CH based on its sensing population, which is defined as the number of sensor nodes that are located within its sensing range. Once a sensor node becomes a CH, it uses its sensing population information to put some nodes into sleep mode in order to save their energy. To do so, a CH selects a random number of sensors to put to sleep. This number should not exceed a specific threshold which is determined by the CH. However, the absence of redundancy check in this process leads to potential coverage holes. %###[notes from Wei: what is the definition of "included"? It is not clean in this context. ] I rewrote the sentence, Done.

Another distributed Coverage-Preserving Clustering Protocol (CPCP) is proposed in \cite{Soro2009955}. The CPCP defines several cost metrics that combine the remaining energy of a node with its contribution to network coverage. For example, the minimum-weight coverage cost metric is defined such that nodes deployed in densely populated network areas and that has higher remaining energy are better candidates to act as CHs and/or to stay active. The operation of CPCP consists of five phases. In the first phase, the sensors exchange information about their remaining energy and each node calculates its coverage cost based on that information.  In the second phase, each sensor decides whether or not to become a CH for the current round based on its activation time. Every sensor determines its activation time based on its current coverage cost. A sensor that does not hear an announcement message, from any other sensor node, during its activation time will declare itself to be a new CH upon the expiration of its activation time. In order to avoid creating non-balanced clusters, a sensor node announce its role as a CH within a prespecified cluster range. In the third phase, a multi-hop route between the CHs and the BS is constructed. In the fourth phase, the clusters are formed such that each non-CH node joins the closest CH. In the final phase, each sensor decides whether it will stay active or not for the current round. This decision is based on its coverage cost. In order to take this decision, every node defines an activation time based on its current coverage cost. Doing that will allow sensors that have lower coverage cost to announce themself earlier as active nodes. Every node will determine its status upon the end of its activation time. If a sensor node determines that its sensing area is completely covered by its neighboring nodes, it turns itself off for the current round. However, this activation method is not efficient, as it cannot gurantee to find all redundant nodes in each round. Moreover, the main operation of CPCP depends mainly on the values of the activation timers. So the decision of whether a sensor will stay active or not is not taken at the begining of the round. This decision could be taken by the node anytime during the round, depending on its activation time. This will lead to unnecessary consumed energy by the redundant nodes  who are waiting their timer to expire to take the decision to be inactive. Moreover, although the authors of CPCP recommended the activation time to be proportional to the coverage cost, no specific recommendation was given on how to set this value.

The authors in \cite{Wang2009838} developed a centralized, cluster-based coverage-aware protocol for target tracking applications. The network area consists of both randomly deployed static sensor nodes and mobile sensor nodes. The main idea of the protocol is to optimize the positions of the mobile sensor nodes to increase the coverage rate of the ROI. Static nodes are partitioned into clusters using maximum entropy clustering. CHs are assigned by placing mobile sensor nodes in the positions of the clusters' center. For each cluster, two metrics are calculated: the coverage metric and the energy metric in order to assess the coverage rate and the energy efficiency, respectively. The coverage metric is defined as the proportion of the detected area to the whole sensing area.  In order to calculate the coverage metric, the whole sensing area is divided into grids and then the grids are simpilified to points. For each point, the number of nodes by which this point is covered is calculated. %###and it is calculated using a gridding %###[notes from Wei: what is gridding algorithm??] I rewrote the sentence, Done
While the energy metric is defined as the lowest cost among all possible communication paths that are from each node to its cluster head. It is calculated using Dijkstra's algorithm. Then, each assigned cluster head performs particle swarm optimization to maximize the coverage metric and minimize the energy metric. A weight coefficient between the two metrics is employed to model the trade-off between coverage rate and energy efficiency. It was assumed that all sensor nodes are location-aware and the energy consumption on relocating the mobile sensor nodes is ignored. The boolean sensing model is used to measure the coverage rate. Unlike the aforementioned coverage-aware clustering protocols, this protocol divides the sensor network into fixed number of clusters. The main reason behind this division is to perform parallel coverage optimization for all clusters. Moreover, no TDMA scheduling mechanism is employed for the inter-cluster communication. Instead, the intra-cluster communication is achieved by constructing a routing tree that connects all the sensors inside a cluster. %###[notes from Wei: this work should be compared to [26]. ] %###[notes from Riham: I am not sure if I should do that sice Mobibar [26] is a deployment protocol in which all the sensors are mobile and is for a different  kind of applications, Also according to our classification in this review, Mobibar belongs to a different category of protocols], Plz let me know if you still want me to compare it.  

In \cite{Özdemir2013}, a pareto-based Multi-objective Evolutionary Algorithm (EA)-based on Decomposition (MOEA/D) algorithm is adopted as an optimization tool to design a coverage-aware clustering protocol, which we will refer here as (MOEAD-CCP). Each candidate solution of MOEAD-CCP has a dimension equal to the network size. Each gene of the candidate solution maps the status of the corresponding sensor. A sensor node has four different states: $-1$ to indicate it is a dead sensor, $0$ to indicate it is in inactive mode, $1$ to indicate it is an active node or $2$ to indicate it is a CH. This encoding results in a variable number of CHs and provides a joint solution for both the clustering and coverage problems. Two main objective functions are defined to evaluate each candidate solution. The first one minimizes the total distance between the sensor nodes to their respective CHs and from the CHs to the BS while the second objective minimizes the total number of uncovered nodes. Although MOEAD-CCP provides a joint solution for both the clustering and coverage problems in WSNs, the objective functions are not well-defined. Similar to FCM-3, the first objective function assumes the that energy consumption can be minimized by minimizing the total distance between the sensor nodes and their respective CHs and from the CHs to the BS. The second objective function of MOEAD-CCP minimizes the number of uncovered points without applying any redundancy check and without taking into consideration minimizing the number of active nodes. Moreover, MOEAD-CCP ignores optimizing the clustering problem and focus more on the coverage problem. Although this protocol is built under the assumption of 2D WSNs, it can be easily adopted for 3D WSNs by calculating the distance using the 3D coordinates of each sensor instead of the 2D coordinates.

A  centralized pareto-based multi-objective approach for designing and developing an energy-efficient, scalable, reliable and cluster-based coverage aware network configuration protocol for 3D WSNs was proposed in \cite{8447593}. The main objective of the proposed protocol is to find a joint and simultaneous solution to maintain full connectivity and coverage in 3D WSNs by finding the optimal status (cluster head, active, or inactive/sleep) for each sensor in the network.  The network field is divided into a set of equal cubes. A new linear programming formulation that handles the connectivity and coverage optimization in 3D WSNs is proposed. These two concerns are formulated as a single multi-objective minimization problem. To do so, a new chromosome encoding scheme is proposed. Moreover, the solution aims at minimizing the number of active sensors to be clustered while providing full area coverage at the same time. The proposed formulation considers the following combined properties: energy efficiency, data delivery reliability, scalability, and area coverage. Two different types of Pareto-based algorithms are considered as optimization tools to solve the joint problem of clustering and coverage in 3D WSNs: the Non-dominated Sorting Genetic Algorithm II (NSGA-II) \cite{NSGA2} and the Multi-objective Evolutionary Algorithm (EA)-based on Decomposition (MOEA/D) \cite{4358754}. However, performance results have shown that  NSGA-II outperforms MOEA/D in terms of many optimization criteria. Hence, NSGA-II was adopted as the optimization tool and the protocol was named  NSGA-II based Coverage-aware Clustering Protocol for 3D WSNs (NSGA-CCP-3D). NSGA-CCP-3D ensures full area coverage by ensuring that each cube is covered by at least one sensor. NSGA-CCP-3D also ensure that all the CHs are connected to the BS and hence ensures connectivity.  NSGA-CCP-3D assumed a binary sensing model and a 3D network field. All sensors are assumed to be static, homogenous, and location-aware. A realistic  energy consumption model that is based on the characteristics of the Chipcon CC2420 radio transceiver data sheet is assumed.
Table \ref{rltdcc} shows a comparison between the coverage aware deployment protocols discussed above.

\begin{table*}[htp]
\centering
\caption{Comparison of cluster-based sleep scheduling protocols}
\resizebox{0.8\textwidth}{!}
{
\begin{tabular}{llllllllllll}
\toprule[1.5pt]
 & \textbf{Coverage}& \textbf{Year} &\textbf{Main} & \textbf{Sensing} &\textbf{Location} & &  \multicolumn{4}{l}{\textbf{Protocol Characteristics}}\\
\cmidrule(l){8-11}
\textbf{} & \textbf{Protocol} & \textbf{Published} & \textbf{Goal(s)} & \textbf{Model} & \textbf{Awareness} &  & \textbf{Dist. } & \textbf{Cent.} & \textbf{EC.}\\
\midrule
 & \cite{Wang2009838} & 2009 & Increase coverage rate/Point coverage & Boolean & Yes & &\xmark & \cmark & \cmark \\
 & BCA \cite{Shin2015} & 2015 &   Create equally balanced clusters/Area coverage &   Boolean &  Yes & & \cmark &  \xmark &  \xmark \\
  & CPCP \cite{Soro2009955} & 2009 &   Clustering/Area coverage &   Boolean &  Yes & & \cmark &  \xmark &  \xmark \\
  & ECDC \cite{Gu2014384} & 2014 & Increase coverage rate/Point/Area coverage & Boolean & Yes & &\cmark & \xmark & \xmark \\
  & MOEAD-CCP \cite{Özdemir2013} & 2014 &  Clustering/Full area Coverage & Boolean & Yes & &\xmark & \cmark & \cmark \\
  & NSGA-CCP-3D \cite{8447593} & 2018 & Clustering/Full area Coverage/3D WSN & Boolean & Yes & &\xmark & \cmark & \cmark \\
  \bottomrule[1.5pt]
\end{tabular}
}
\label{rltdcc}
\end{table*}

\section{Open Issues}
\label{OI}
%Several protocols to solve the coverage problem in WSNs were discussed in the previous Section. However, there are still a lot of open issues that need to be resolved. 

Most of the proposed coverage protocols assume nonrealistic assumptions about the underlying system model. In the following paragraphs, we discuss those issues in details. Furthermore, we discuss other relevant issues that affect the performance of the coverage protocols.  

\subsection{Realistic Sensing Model}
%Many coverage protocols have been recently proposed in the literature to solve the coverage problem. However, the performance of these protocols is significantly affected by the assumptions made about the underlying sensing model. 
Majority of the existing coverage protocols use the simple and ideal boolean sensing model. However, it is unlikely that sensing signals drop suddenly from full-strength to zero, as assumed by the boolean sensing model. In reality, the sensing range around a sensor is not uniformly distributed.As a result, one the following problems will likely arise:
\begin{itemize}[noitemsep,nolistsep]
\item The boolean sensing model may not fully represent the sensing capacity of the sensors as there might be case that points located beyond the defined uniform sensing range are already covered. Consequently, network lifetime is decreased because more (redundant) sensors than required are kept active.
\item On the other hand, boolean sensing model may also overestimate the sensor's sensing capacity by assuming that all the points located within its uniform sensing range are covered. This in turn may result in coverage holes in the network and the coverage requirement of the application is not met.

\end{itemize}

The Elfes sensing model \ref{elfes} was proposed to solve those problems by defining two sensing ranges for a sensor. However, the coverage area for each sensing range is still uniformly distributed and hence it shares the same problems as the boolean sensing model. %###[comments from Wei: I do not agree with this sentence. In my opinion, the above-mentioned reason that you have provided is what Elfs model trys to solve. What's your opinioin?] I rewrote this line, Done.

Several empirical studies have shown that the shape of the sensing area of a sensor may not be a regular disk \cite{Dhraief} \cite{4850343} \cite{Zhou} %###[comments from Wei: add references here]. Done
The impact of location errors, sensing signal irregularity and packet loss on the Coverage Configuration Protocol \cite{Xing:2005}, CCP, were studied and investigated in \cite{Dhraief}. Experimental results shown that CCP performance degrades with the location errors increase, sensing signal irregularity and packet losses. Moreover, the impact of radio irregularity on the sensor communication was confirmed and quantified in \cite{4850343}. According to \cite{Zhou}, the radio irregularity in WSNs is caused by three main factors:

\begin{itemize}[noitemsep, nolistsep]
 \item \textbf{Anisotropy:} a signal transmitted by a sensor node experiences various path losses at different directions.
\item \textbf{Continuous variation:} the signal path loss varies continuously with incremental changes of the propagation direction from a transmitter.
\item \textbf{Heterogeneous sending powers:} sensor nodes may transmit radio signals at different sending powers, even though they are from the same manufacturer. This is caused due to the hardware differences between sensors and different battery level of the sensors. 
\end{itemize}

The shadow fading sensing model was proposed as a first attempt to construct a more realistic sensing model by simulating the path loss around the sensors. However, a careful look at the coverage protocols in the literature will reveal that this model has rarely been used to model the coverage of the sensor in WSNs. Moreover, the shadow fading sensing model is isotropic in the sense that the path losses in different directions are the same \cite{Zhou}. To illustrate this, the path loss at distance $d$, $PL(d)$ is calculated using the following equation \cite{Dezfouli2015102}:

\begin{equation}
PL(d) = PL(d_0) + 10\beta \log_{10} (\frac{d}{d_0}) + N(0,\sigma)
\label{pl}
\end{equation}

Where $d$ is the distance from the sender, $d_0$ is the reference distance, $PL(d_0)$ is the path loss at a reference distance $d_0$, $\beta$ is the path-loss exponent, $N(0,\sigma)$ is a zero-mean Gaussian random variable with standard deviation $\sigma$.

The Radio Irregularity Model (RIM) was proposed in \cite{Zhou} to simulate the three factors that cause radio irregularity in WSNs. In RIM, the irregularity of a radio pattern is denoted by parameter Degree of Irregularity, DOI. DOI is defined as the maximum path loss percentage variation per unit degree change in the direction of radio propagation. Accordingly, the path loss model is modified based on the DOI to generate $360$ different path loss values for different directions. In RIM, the DOI-adjusted path loss at direction $\theta$ and distance $d$, $PL(d, \theta)$ is calculated using the following equation \cite{Dezfouli2015102}:

\begin{equation}
PL(d, \theta) = (PL(d_0) + 10\beta \log_{10} (\frac{d}{d_0})) \times K_{\theta} + N(0,\sigma) 
\label{plrim}
\end{equation}

Where $K_{\theta}$ is the path-loss coefficient at direction $\theta$ and is computed as follows:

\begin{equation}
K_{\theta} = \begin{cases}
    \text{if }\theta = 0: K_{\theta} = 1\\
    \text{if }{0 < \theta < 360}: \begin{cases}
    \text{if }B(1,0.5)= 0:\\ K_{\theta} = K_{\theta-1} + W(b_1,b_2) \times DOI\\
    \text{if }B(1,0.5)= 1:\\ K_{\theta} = K_{\theta-1} - W(b_1,b_2) \times DOI\\
\end{cases}
\end{cases}
\label{k}
\end{equation}

Where $|K_{0} - K_{395}| < DOI$, $B(n, p)$ is a Binomial random variable with $n$ trials and success probability $p$, and $W(b_1,b_2)$ is a Weibull random variable with scale parameter $b_1$ and shape parameter $b_2$. 

RIM is a thoughtfully designed model that reflects the signal irregularity phenomena in WSNs \cite{Liu201568} \cite{Dezfouli2015102}. It has been used and studied recently in many WSNs protocols such as \cite{Phoemphon2016} \cite{Kumar2015438} \cite{7545125} \cite{6214058} \cite{Dezfouli2015102}. However, most of these protocols focus on localization methods or link quality estimation. To the best of our knowledge, RIM has not yet been studied or used to model the coverage problem in WSNs.

\subsection{Realistic Coverage-aware Clustering Protocols}
Clustering sensor nodes is an efficient topology control method to maximize the network's energy efficiency. Many clustering protocols have been used in various WSNs applications. However, most of these protocols focus only on selecting the optimal set of CHs to reduce or balance the energy consumption of a given network, while how to cover the network area effectively is overlooked. Moreover, the performance of these protocols is limited by the challenges on determining an accurate radio model for the sensor nodes in the network. A commonly employed energy consumption model is presented in \cite{real1} \cite{real2}. The energy consumption in this model is calculated as follows:

\begin{equation}
E_{TX}(k,d) = \begin{cases}
(E_{elec} + \varepsilon_{fs} \times d^2) \times k, d \leq d_{0}\\
(E_{elec} + \varepsilon_{mp} \times d^4) \times k, d > d_{0}
\end{cases}
\end{equation}

\begin{equation}
E_{RX}(k) = E_{elec} \times k
\end{equation}

Where $E_{elec}$ stands for the energy consumption required to run the transmitter or the receiver circuitry. $d_{0}$ is the distance threshold. $\varepsilon_{fs}$ and $\varepsilon_{mp}$ are the required energies for amplification of transmitted signals in the open space and the multi-path models respectively.

However, this energy model presented in \cite{real1} \cite{real2} is idealized and fundamentally flawed for modelling radio power consumption in sensor networks. It ignores the listening energy consumption, which is known to be the largest factor to expend energy in WSNs \cite{real1} \cite{real2} \cite{real3}. Moreover, it assumes that the communication range between any pair of sensor nodes is infinite. 

A discrete radio model should be used for more accurate and realistic calculation of the power consumption and to determine which links between sensor nodes are available for transmission \cite{real1} \cite{real2} \cite{real3}. A realistic energy consumption model based on the characteristics of the Chipcon CC2420 radio transceiver data sheet \cite{CC2420} can be used to model the energy consumption in coverage protocols. The total energy consumed by node $ i $, $ E_{i} $, is calculated as follows \cite{radio-use}:

\begin{equation}
E_{i}=\displaystyle\sum_{state j}P_{state j}\times t_{state j}+\sum E_{transitions} \hspace{10 mm}
\end{equation}

The index $state j$ refers to the energy states of the sensor: sleep, reception, or transmission. $  P_{state j} $ is the power consumed in each $ state j $, and $ t _{state j} $ is the time spent in the corresponding state. Moreover, the energy spent in transitions between states, $ E_{transitions} $, is also added to the node's total energy consumption. The different values of $  P_{state j} $ and  $ E_{transitions} $ can be found in \cite{CC2420}.

Although recent efforts have been made to develop clustering protocols under realistic energy consumption models such as \cite{Elhabyan2015116} \cite{7237275} \cite{7051910}, further enhancement on energy-efficiency may be obtained if these solutions have adopted a sleep-scheduling mechanism before or after the clustering process. Furthermore, to enhance the network's energy efficiency, integrated solutions that are based on realistic energy as well as sensing models to address coverage problems using clustering approaches should be investigated.

\subsection{Realistic Connectivity Model}
As discussed earlier in Section \ref{cvrgrltd}, both coverage and connectivity problems should be jointly investigated \cite{Zhao20082205} \cite{Goel:2014} \cite{Ghosh2008303}. Indeed, there are many previously proposed protocols that consider both of these problems simultaneously \cite{7217790} \cite{Khalesian2016126} \cite{Gupta2015} \cite{Xing:2005}. However, these protocols use the distance between two sensors as the metric to evaluate a link quality. Moreover, most of these protocols assumed that the network connectivity will be guaranteed if full area coverage is achieved and the communication range is at least twice the sensing range \cite{Ghosh2008303} \cite{Wang:2003}. However, these assumptions suffer from the following two problems:

\begin{itemize}[noitemsep, nolistsep]
\item In order to calculate the distance between two nodes, each node should be equipped with either a locating aware device such as a global positioning system (GPS) or a distance measuring device, which in turn, renders a more expensive solution \cite{ld}.

\item Link asymmetry is an important characteristic of WSNs. Using the distance between two nodes as a link quality metric fails to consider this fact.

\end{itemize}

Several studies have shown that link quality in WSNs is not necessarily correlated with distance \cite{rssi2010} \cite{rssi} \cite{rssi06} \cite{rssi2006}.

Two other prominent link-quality metrics are the Received Signal Strength Indicator (RSSI) and Link Quality Indicator (LQI). These metrics are provided by most of the wireless sensor chips \cite{6879479}. The RSSI is a parameter that represents the signal strength observed at the receiver at the moment of reception of the packet. The LQI is described as the characterization of the strength and quality of the received packets.

Several studies prove that RSSI can provide a quick and accurate estimate of whether a link is of very good quality \cite{rssi} \cite{rssi06} \cite{rssi2006} \cite{rssi2010}. In \cite{rssi2006}, the authors conducted empirical measurements of the packet delivery performance of various sensor platforms. They found that there was a strong correlation between RSSI and Packet Delivery Rate (PDR). Furthermore, they proved that if the RSSI of a link is -$ 87dBm $ or stronger, it is almost but not completely set to achieve a $ \ge 99\% $ PDR. Below this value, a shift in the RSSI as small as 2 dBm can change a good link to a bad one and vice versa, which means that a link is temporality located in a transitional or disconnected region \cite{rssi2010}.

The symmetry of RSSI and LQI in two directions was studied, and the relation between RSSI and LQI as link quality metrics is analyzed in \cite{6879479}. Experimental results show a significant correlation between the two directions of the link in RSSI but a weak correlation between the two in LQI. Moreover, statistical tests on the collected data show a significant correlation between RSSI and distance in short distance scenarios, which makes RSSI a routing protocol link-quality metric. Therefore, the RSSI value should be used to assess the quality of a link between any pair of sensors.

%###[comments from Wei: i have not edited line 587- 623] done
\subsection{Sensors Localization}

Sensors locations is another fundamental problem in WSNs. Location information of sensors is usually a key input to solve the coverage problem. Many of the previously proposed coverage protocols simplified their solution by assuming that each sensor is location aware given it is equipped with a self-locating hardware such as a GPS. Though this assumption allows the design of simple and efficient solutions, the resulting costs render such solutions ineffective and unrealistic \cite{ld}. In order to design more cost effective coverage protocols, no assumption should be made on the location awareness of sensor nodes in the WSNs. That is, alternative solutions using sensors without GPS is consequently highly required.

Many localization protocols have been proposed to enable stationary \cite{5191633} or mobile sensors \cite{mlc} \cite{5325242} to autonomously determine their positions without relying on GPS. Localization protocols can be classified into either range-based protocols or range-free protocols. Range-based protocols assume that the sensors are equipped with a specific measurement device to measure the distance or the angle between themselves and other regular sensors with unknown locations \cite{He}. Unlike range-based protocols, range-free localization protocols rely on the network connectivity to estimate the sensors positions. Due to this reason, range-free localization protocols have recently received the attention of the research community. Many range-free localization protocols have been proposed in the literature. However, most of them do not consider the anisotropic nature of WSNs \cite{7524770} \cite{Zhang:2015}.

Despite the close relationship between the coverage problem and the localization problem, they have been discussed and evaluated separately. It is beneficial to design protocols that solve the localization problem and the coverage problem under the same settings consistently. For example, range-free localization protocols adopted for anisotropic WSNs should be followed by a realistic coverage control protocol.

\subsection{Transmission Power Control}
A method to significantly reduce the energy consumption in WSNs is to apply Transmission Power Control (TCP) techniques to adjust the transmission power \cite{7808888} \cite{Rault2014104} of sensors dynamically. Traditionally, sensor nodes transmit packets at the same power level that is normally the maximum possible power level. However, a node transmitting packets at the highest power level generates too much interference %###[notes from Wei: add explanation or reference here.] Done
in the network and consume more energy than necessary \cite{6339119} \cite{Akbari}. In the case of communication between two nodes that are close to each other, a low transmission power is sufficient. This power level should be high enough just to guarantee the connectivity and low enough to save energy and minimize network interference.
Although both coverage problem and TCP techniques have been studied extensively separately, only a few protocols considered them in a joint way such as \cite{Xenakis2016576} \cite{Wang2014212} \cite{Maurya201620}. However, most of these protocols assumed that all sensors are equally equipped with the same transmission range and sensing range. Our intuition is that relaxing such assumptions will affect the protocols' performance. Therefore, embedding TPC techniques into existing coverage protocols in order to develop more energy-efficient coverage protocols should be investigated.

\subsection{Coverage Protocols using Evolutionary Computation}
As analyzed in detail in Sections \ref{CADP} and \ref{SSCP} of this paper, there are many EC-based protocols that have been proposed to solve the coverage problem in WSNs. However, developing solutions with high-performance EA for the coverage problem in WSNs remains an open issue.

\subsubsection{Individual/Solution Representation}
One of the basic features of EA is that it maintains a population of \textit{individuals} that represent possible solutions to a problem \cite{WICS:WICS5}.
%###[notes from Wei: you should briefly explain "individual representation". Not all researchers reading this paper know about AI or other optimization techniques. ] Done
One of the critical issues in designing efficient %###[notes from Wei: in terms of what else other than time?]
 EA-based protocols  is how to %###[notes from Wei: I have deleted "represent and encode". Instead I used "evaluate". Please make sure this is what you want to express.]Not the same, Done
 represent and encode, more efficiently, each potential individual/solution in the population. Most of the proposed EA-based coverage protocols start with random initialization which may result in many infeasible solutions that may not satisfy some of the problem constraints, such as the connectivity to the BS \cite{7394103}. This, in turn, will affect the search performance and may lead to solutions that achieve less than optimal results. Therefore, there is a need to find new individual/solution representation schemes.%###[notes from Wei: check on this previous sentence too.] Done

\subsubsection{Parameter Settings}
How to determine the parameter settings for the adopted EA is another important issue. Most EA-based protocols assumed default values for setting the parameters of their adopted EA. That is, the impact of parameter setting may have on the results is often ignored. Results show that even running the same protocol may produce results that are quite different under different parameter settings \cite{7394103}. For example, in order to solve the coverage problem in an increased dimension (i.e. increased network size) %###[notes from Wei: 1. what "problem" do you want to say here, coverage problem? 2.  "(increasing has been changed into) increased dimension": what do you mean? The total size of the sensors in the network or something else?], Rewrote it, Done
it is necessary to increase the population size as well as the number of iterations executed by the EA.

\subsubsection{Improved Metaheuristic Approaches}
Many EAs face challenges such as poor exploitation, and slow convergence rate. This explains the increasing need for adopting hybrid and improved EA \cite{Yi20132433}. Improved solution search equation should be adopted to improve the exploitation capabilities of different EAs.

\subsubsection{Multi-objective Approach}
It should also be noted that most of the EA-based protocols use the weighted-sum approach when formulating the multiple objectives of the coverage problem in WSNs. Using the conventional weighted-sum approach in multi-objective optimization is computationally efficient and straightforward to implement \cite{Konak} \cite{Liu} \cite{Lam}. It has been widely used because of its simplicity. However, it is known that this approach has the following problems \cite{Zhao2014407} \cite{Gobbi} \cite{Chiandussi2012912} \cite{Anne}:
\begin{itemize}[noitemsep, nolistsep]
\item This approach results in only one optimal solution. %###[notes from Wei: do you mean "a solution achieving optimal solution to only one objective can be obtained from each single run"?] I rewrote it, Done
\item This approach can not find the optimal solution when the feasible solution set in the objectives domain is not convex.
\item The choice of the weight vector can highly affect %###[maybe you want to use "bias" instead?] Done
 the obtained solutions and make them more biased towards one sub-objective than another. 
\end{itemize}
These problems are particularly critical if the objectives are conflicting or must be handled simultaneously. In those cases, the concept of the optimal solution changes because the end goal becomes to select one from a set of good trade-off solutions. Such selection is usually difficult and done manually by the decision makers.

Moreover, many other WSNs problems such as deployment, localization, clustering, and connectivity are also formulated as optimization problems. A solution that simultaneously solves all or more than one of such problems is a multi-objective task. Hence, we define the \textit{WSNs lifetime maximization problem} as a comprehensive problem of minimizing the energy consumption in all sensor network stages, i.e, deployment, localization, clustering, coverage, and connectivity. A multi-objective optimization framework should be developed to solve this problem. However, finding an individual encoding which reflect a joint solution for all these problems is a challenging task \cite{7570253}. In order to apply the multi-objective approach in the WSNs lifetime maximization problem, a new individual encoding scheme which represents a joint solution for all the above-mentioned problems should be studied.   %###[notes from Wei: this should be further explained and referenced] Done

\section{Conclusions
%and Future Directions
}\label{CFD}  %###[comments from Wei: supposedly open issue Section already included all open Issue and future directions in the field. Since this is not a technical paper, there is no point adding future directions in the conclusion Section. ] Done
Recently, many new solutions have been proposed to solve the coverage problems in WSNs. In this survey, we presented a thorough and up to date review of these coverage protocols. We found that the performance of these protocols is mainly limited by challenges related to determining a more realistic coverage model for the sensor nodes in the networks. More specifically, most of the proposed coverage protocols rest on less realistic assumptions such as location awareness and uniformity of the signal strengths within a sensing and/or a communication range. Furthermore, most of these protocols use an idealized energy consumption model. We believe that a discrete radio model should be used to achieve a more accurate and realistic calculation of the power consumption to enable a better link selection for transmission.

Motivated by the aforementioned concerns, we carefully study, compare and analyze in detail all known coverage protocols on different design factors/features. Finally, we point out open problems and future research directions such as addressing coverage problem in a more realistic sensing model that reflects the anisotropic properties of WSNs. Most importantly, we conclude that network connectivity is a crucial factor that must be taken into consideration in designing future solutions. 

\section{References}
\bibliographystyle{ieeetr}
\bibliography{references}

\epsfysize=3.2cm
\begin{biography}{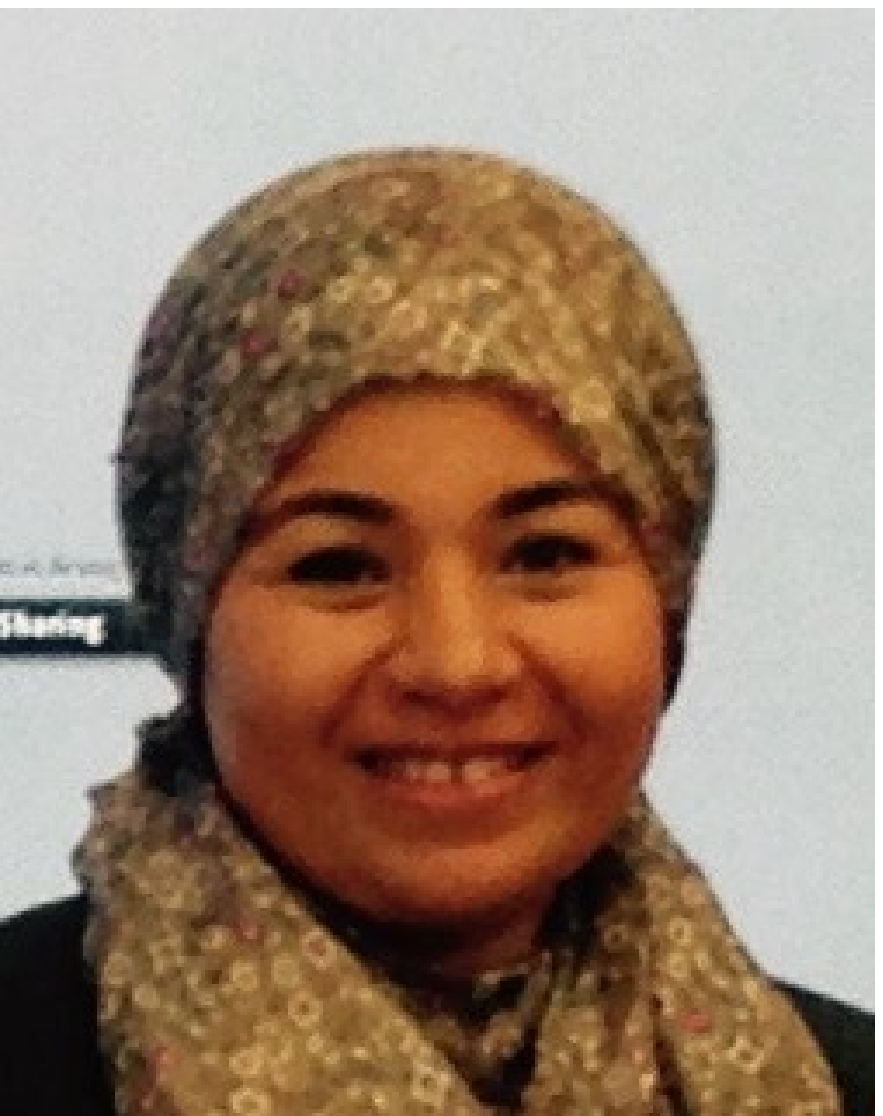}{Riham Elhabyan} Dr. Riham Elhabyan currently works as a Database Analyst with the Department of Fisheries and Oceans, Canada.  Previously, she has worked as a Postdoctoral Fellow with the School of Information Technology at Carleton University. Dr. Elhabyan is specialized in applying Artificial Intelligence approaches to solve real world problems in areas such as Big Data Infrastructures and Wireless Communications. She holds a Ph.D. (Artificial Intelligence) from the University of Ottawa, Canada and a M.SC. (Distributed Computing) from the University of Nottingham, UK. Recently, Dr. Elhabyan has been awarded the 3d prize in the IEEE Women in Engineering (WIE) Best Paper competition.
\end{biography}
\epsfysize=3.2cm
\begin{biography}{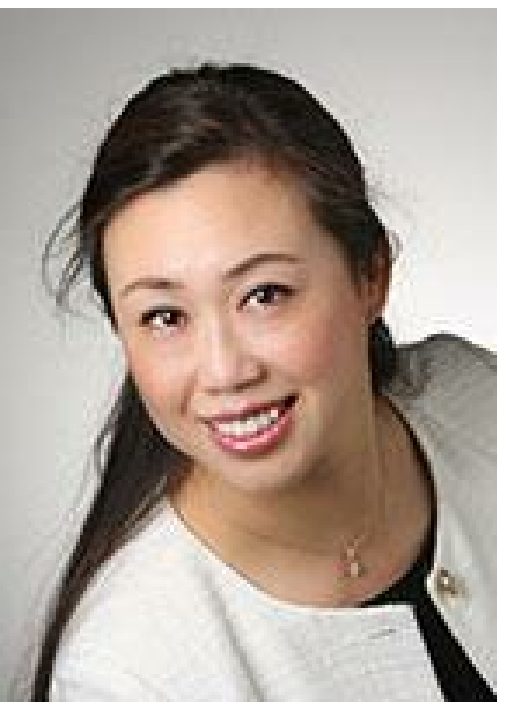}{Wei Shi} Dr. Wei Shi is an Associate professor in the School of Information Technology, cross-appointed to the Department of Systems and Computer Engineering in the Faculty of Engineering and Design at Carleton University. She specializes in algorithm design and analysis in distributed environments such as Data Centers, Clouds, Mobile Agents and Actuator systems and Wireless Sensor Networks. She has also been conducting research in data privacy and Big Data analytics. She holds a Bachelor of Computer Engineering from Harbin Institute of Technology in China and received her Master's and Ph.D. of Computer Science from Carleton University in Ottawa, Canada. Dr. Shi has published over 60 articles in reputable conferences and journals. She is also a Professional Engineer licensed in Ontario, Canada.

\end{biography}
\epsfysize=3.2cm
\begin{biography}{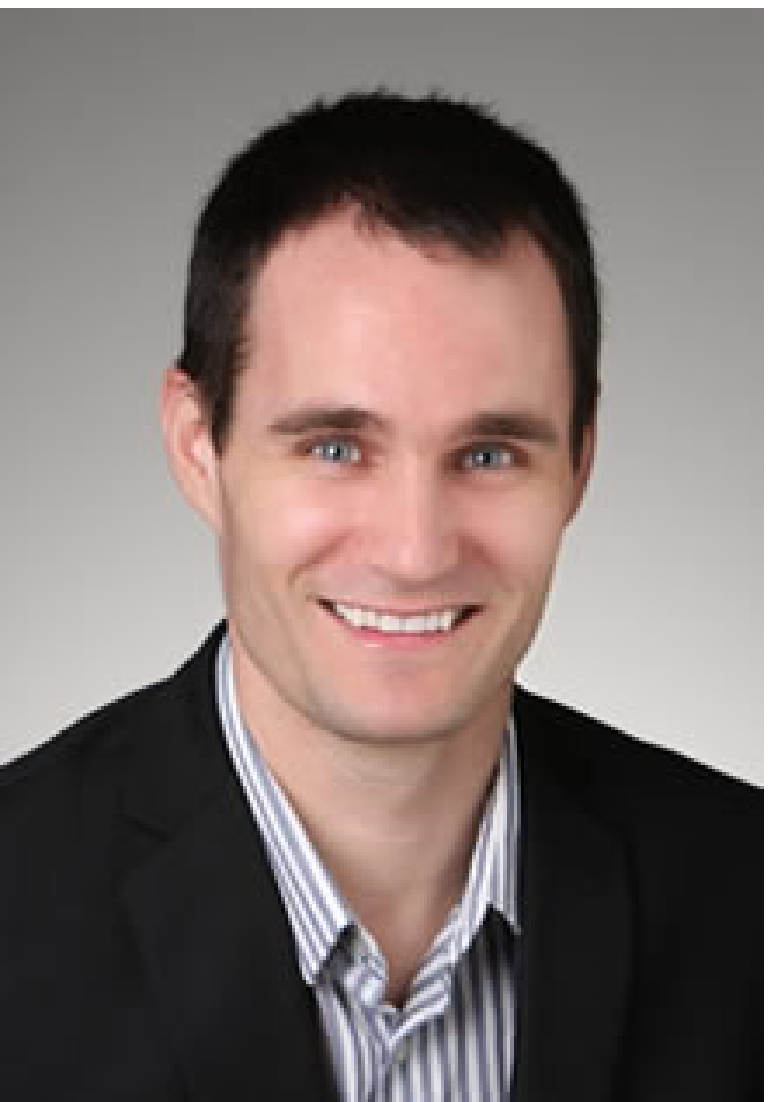}{Marc St-Hilaire} Marc St-Hilaire joined Carleton University in 2006 upon completion of his PhD in Computer Engineering, from École Polytechnique of Montreal. He is currently an associate professor with the School of Information Technology with a cross appointment with the Department of Systems and Computer Engineering at Carleton University. Dr. St-Hilaire is conducting research on various aspects of wireline and wireless communication systems. More precisely, he is interested in network planning and infrastructure, network protocols, network interconnection, and performance analysis. With more than 130 publications, his work has been published in several journals and international conferences. Finally, Dr. St-Hilaire is actively involved in the research community. In addition to serving as a member of technical program committees of various conferences, he is equally involved in the organization of several national and international conferences and workshops. He is also a senior member of the IEEE.
\end{biography}
\end{document}